\documentclass[aps,twocolumn,showpacs,amsmath,amssymb]{revtex4}

\usepackage{graphicx}

\begin{document}

\title{Interaction of carbon clusters with Ni(100) : \\
Application to the nucleation of carbon nanotubes}

\author{H. Amara}
\affiliation{Laboratoire de Physique du Solide, 
Facult\'es Universitaires Notre-Dame
de la Paix, \\ 61 Rue de Bruxelles, 5000 Namur, Belgique}
\affiliation{PCPM $\&$ CERMIN, Universit\'e Catholique de Louvain,
 Place Croix du Sud 1, 1348 Louvain-la Neuve, Belgique} 
 \altaffiliation[Present address: ]{Laboratoire
Francis Perrin, CEA-CNRS, B\^{a}timent 522,
91191 Gif sur Yvette Cedex, France}
\email{hakim.amara@cea.fr}
\author{C. Bichara}
\affiliation{CRMCN-CNRS, Campus de Luminy, 
Case 913, 13288 Marseille Cedex 09, France} 
\author{F. Ducastelle}
\affiliation{Laboratoire d'Etudes des Microstructures, 
ONERA-CNRS, BP 72, 92322 
Ch\^atillon Cedex, France}

\begin{abstract}
In order to understand the first stages of the nucleation of carbon nanotubes in catalytic processes, we present a tight-binding Monte Carlo study of the stability and cohesive mechanisms of different carbon structures deposited on nickel (100) surfaces. Depending on the geometry, we obtain contrasted results. On the one hand, the analysis of the local energy distributions of flat carbon sheets, demonstrate that dangling bonds remain unsaturated in spite of the presence of the metallic catalyst. Their adhesion results from the energy gain of the surface Ni atoms located below the carbon nanostructure. On the other hand, carbon caps are stabilized by the presence of carbon atoms occupying the hollow sites of the fcc nickel structure suggesting the saturation of the dangling bonds.
\end{abstract}


\maketitle

\section{Introduction} 

During the past few years, there has been a tremendous increase in research activities devoted to single-wall carbon nanotubes (SWNTs) mainly because of their extraordinary mechanical, electrical, and chemical properties. Future nanoscale applications will however necessitate the production of pure SWNTs in significant quantities and also the selective formation of nanotubes with well-defined chirality or diameter. This obviously requires to understand their growth mechanisms. SWNTs are generally produced via techniques working at high temperatures (arc discharge \cite{Journet1997}, laser ablation \cite{Thess1996}) or at medium temperature (chemical vapor deposition \cite{Colomer1999}). The presence of a transition metal catalyst (such as Fe, Co, Ni, $\dots$) has been found necessary for the synthesis of SWNTs but the exact role played by the catalyst is still unclear. The similarities between the samples synthesized from different techniques suggest a common growth mechanism and the so-called vapor-liquid-solid (VLS) model is frequently mentioned \cite{Gavillet2001,Gavillet2004}. In this scheme, the growth of SWNTs begins with the formation of a metallic particle supersaturated with carbon; then carbon segregates towards the surface and finally carbonaceous structures emerge from the catalyst surface. From the theoretical side, several modelizations at an atomic scale have been performed using either phenomenological potentials \cite{Maiti1995,Maiti1997,Shibuta2003,Ding2004} or first principles methods \cite{Gavillet2001,Fan2003,Raty2005,Reich2005}.

To some extent these computer simulation works confirm the VLS model but with some differences between the high temperature and medium temperature routes. In chemical vapor deposition methods the size of the tubes is generally determined by the size of the particle. Actually several modelizations show that carbon atoms at the surface self-organize and form graphene-like sheets wrapping the particle. Once formed these sheets no longer interact with metallic atoms and can detach to form nanotubes. This accounts for a so-called tangential process. In other situations, principally within the high temperature route, bundles of nanotubes grow perpendicularly to the surface and it is still a challenge to understand why this can be more advantageous than a tangential growth.

Some arguments based on classical nucleation and growth thermodynamic models have already been put forward  \cite{Kuznetsov2001,Kanzow2001}. Carbon atoms at the surface of the metallic catalyst are assumed to condense in the form of graphene flakes. Then the metallic substrate can help to saturate the dangling bonds and this is presumably favoured by the formation of a cap, the energy cost due to the curvature induced by the presence of pentagons, being more than compensated by the reduction of the number of dangling bonds \cite{Kanzow2001}. This model has been substantiated by Fan \textit{et al.} \cite{Fan2003} who performed \textit{ab initio} energy calculations of different arrangements of carbon atoms on a Ni(100) surface. Among the different carbon geometries tested, they found actually that the most stable one is a hemispherical cap forming a tube embryo. These calculations are quite accurate but computationally very demanding and the structures cannot be fully relaxed. Furthermore these calculations do not provide an easy access to a local analysis of the bonding. Fan \textit{et al.} have thus developed a simplified phenomenological analysis of the energies to characterize this effect of dangling bond saturation~\cite{note}. A more precise analysis of the chirality fixed by the embryos growing on a (111) surface has been made by Reich \textit{et al.} \cite{Reich2005} based on the use of the SIESTA \textit{ab initio} code. They confirm the bond saturation effect and show that the possible chiral selectivity depends more on the carbon-metal bonds than on the energy of the different carbon caps. In this calculation however, the metallic (nickel) atoms remain unrelaxed, a strong approximation, considering the order of magnitude of the NiC interaction, and no local energy analysis can either be achieved.

In this paper, we study the adhesive properties of carbon clusters on a Ni(100) surface using a tight-binding (TB) total energy model. This model relies on local (atomic) energy calculations using the recursion method and is coupled with Monte Carlo (MC) simulations in order to relax the structures.  Our objective is to establish if the nucleation of SWNTs is favoured by the elimination of dangling bonds, as suggested by the above mentioned authors and also to discuss the role of the displacements of the metallic atoms. We thus present an analysis of the local energy distributions
in order to determine which atoms (carbon or nickel), and to what extent, are stabilized when various carbon clusters are put in contact with a metallic surface.

\section{Method} 

In our study, the interaction between metal (here nickel) and carbon atoms is treated within the semi-empirical tight-binding model described in Ref. \cite{Amara1}. Only \textit{s}, \textit{p} electrons
of carbon and \textit{d} electrons of nickel are taken into account. The total energy of an atom $i$ is splitted in two parts, a band structure term that describes the formation of an energy band when atoms are put together and a repulsive term that empirically accounts for the ionic and electronic repulsions :
\begin{equation}
E_{tot}^{i} = E_{band}^{i} + E_{rep}^{i}.
\label{eqn:equation1}
\end{equation} 
The total energy of the system, $E_{tot}$, then writes :
\begin{equation}
E_{tot} =  \sum_{{\rm i \ atoms}} E_{tot}^{i}.
\label{eqn:equation2}
\end{equation} 
The band energy, $E_{band}^{i}$, is given by : 
\begin{equation}
E_{band}^{i} =  \int_{-\infty}^{E_f}  (E - \epsilon_i^{0}) n_i(E) dE \mbox{,}
\label{eqn:equation3}
\end{equation} 
where $E_f$ denotes the Fermi level and $\epsilon_i^{0}$ is the atomic energy level. We use the recursion method to calculate the local electronic density of states $n_{i}(E)$ on each site \cite{Haydock}. Only the first four continued fraction coefficients, ($a_{1}$, $b_{1}$, $a_{2}$, $b_{2}$) corresponding to the first four moments of the local density of states are calculated exactly. The continued fraction is then expanded to the $N^{th}$ using constant coefficient equal to $a_{2}$ and $b_{2}$ (here $N=40$). The local density of states $n_{i}(E)$ is then expressed as a set of $N$ poles ($E_{i}^{j}$, $j=1,N$) with weights ($A_{i}^{j}$, $j=1,N$) by diagonalizing a tridiagonal matrix of size $N$. Thus, the band energy of an atom $i$ writes :
\begin{equation}
E_{band}^{i} =  \sum_{j=1}^{j_{max}}A_{i}^{j}E_{i}^{j}
\label{eqn:equation4}
\end{equation}
In the above equation the highest occupied energy level $j_{max}$ depends on each site $i$. The set of C--C, Ni--Ni and Ni--C tight-binding interaction parameters used was shown to be transferable to different atomic configurations by comparison with \textit{ab initio} calculations. It reproduces the expected phase separation tendency for the Ni-C alloy, with an enthalpy of mixing of +1.0 eV/atom for NiC in the (metastable) rocksalt structure fitted to \textit{ab initio} data (ABINIT code). The heat
of solution of a C atom in a Ni matrix is then found equal to +0.4 eV/atom, in agreement with the experimental value of Shelton \textit{et al.} (0.49 eV/atom \cite{Blakely1974}) and the first principles calculations of Siegel \textit{et al.} (+0.2-0.35 eV/atom \cite{Siegel2003}). More details are given elsewhere~\cite{Amara1}. This approach which correctly describes the chemical bonds of a mixed
metal-carbon system, is able to deal with systems of a few hundreds of atoms over large time scales. As compared to the standard diagonalization technique, the recursion method is local and then
gives direct access to the total energy of each atom $i$ (see Eq.\ref{eqn:equation4}).  This possibility to easily analyze local energy distributions will be very useful in the following.

\begin{figure}[htbp!]   
\begin{center}
\includegraphics[width=8cm]{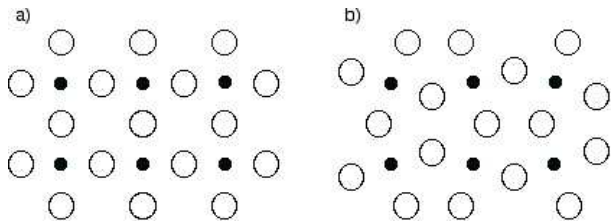}
\end{center}
\caption{(a) Top view of a surface coverage equal to 0.5 ML. (b) Clock reconstruction leading to a $(2\times2)p4g$ symmetry.}
\label{FigClock}
\end{figure}

This atomic interaction model is then implemented in a Monte Carlo code, based on the Metropolis algorithm, which allows to relax the structures \cite{Metropolis1953}. Finally different tests were performed to check the reliability of our model. As an example, a simple relaxation of C atoms covering a Ni(100) surface induced the ``clock" reconstruction observed experimentally by Klink \textit{et al.} \cite{Klink1993}. This reconstruction has also been studied theoretically \cite{Alfe1999,Stolbov2005} and corresponds to the rotation of the four Ni atoms  surrounding a C atom in an Ni(100) surface, clockwise and counterclockwise alternatively (see Fig.\  \ref{FigClock})

In a previous study, we presented Grand Canonical Monte Carlo simulations of the dissolution and segregation of C in the presence of a Ni(111) surface \cite{Amara2} in order to understand the processes involved in the catalytic growth of carbon nanotubes.  At finite temperature (1000 and 1500 K), we analyzed the self-organization of carbon atoms close to a nickel surface and identified a succession of four types of configurations, corresponding to an increase of the carbon chemical potential (and consequently an increase of the number of carbon atoms): single C atoms adsorbed on the surface or incorporated in interstitial sites, chains creeping on the surface, detached $sp^{2}$ C layers and finally a 3D amorphous C phase.
\begin{figure}[htbp!]  
\begin{center}
\includegraphics[width=8cm]{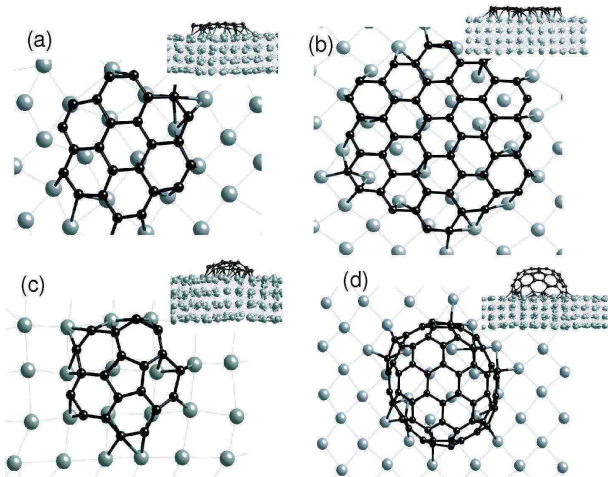}
\end{center}
\caption{Stability at 500 K of carbon clusters (top and side views) deposited on Ni(100). (a) flake of 24 atoms containing 7 hexagons; (b) flake of 54 atoms containing 19 hexagons; (c) curvature of the flake of 20 atoms containing a pentagon; (d) embryo of (6,6) nanotube.}
\label{FigClusters}
\end{figure}

\section{Results} 
Here, we use Monte Carlo simulations in the canonical ensemble (with a fixed number of Ni and C atoms) to relax the atomic structures and analyze the stability of specific carbon structures deposited on a nickel surface. We perform MC simulations on a Ni slab presenting a (100) surface. This is a four-layer slab of 288 atoms, with bond lengths of 2.48 \AA, separated by a 20 \AA \ thick vacuum region along the $z$ axis and periodic boundary conditions. The slab size is 17.60 $\times$ 19.36 $\times$ 5.28 \AA$^{3}$. We consider several configurations of carbon atoms, with bond lengths of 1.42 \AA,
deposited on the metallic surface: molecules (C$_{5}$ and C$_{6}$ rings), a finite linear chain of 12 atoms parallel to the surface, flat flakes of 24 and 54 atoms (see Fig.\ \ref{FigClusters}a--b) and curved structures such as a flake of 20 atoms containing a pentagon (see Fig.\ \ref{FigClusters}c) and the cap of a (6, 6) tube (see Fig.\ \ref{FigClusters}d).
 
First, we study the stability of these carbon clusters in the presence of the metallic substrate at 500 K. At this temperature, the C$_{5}$ and C$_{6}$ rings are dissociated to present C-C bonds of roughly 2.2\AA \ and C-Ni bonds of 1.9 \AA. The twofold coordinated atoms belonging to the finite linear chain interact moderately with the metallic surface (in the 1.0 eV/atom range), while we observe that end of chain, one-fold coordinated, C atom tends to penetrate the metallic substrate, to occupy a semi-octahedral site of the Ni(100) surface. This process has also been observed experimentally by Fujii \textit{et al} \cite{Fujii06}. The stability of the clusters seems to depend on the number of of C-C bonds, with a large enough number of such bonds resulting in a stable cluster. For small number of C-C bonds, the competition between the formation of a carbide and the phase separation results in the formation of a Ni-C alloy. For the other structures, the number of C-C bonds is large enough to stabilize the carbon clusters despite the presence of the metallic surface. The flakes of 24 (Fig.\ \ref{FigClusters}a) and 54 (Fig.\ \ref{FigClusters}b) atoms remain in plane and we also do not observe any dissociation for the flake of 20 atoms containing a pentagon (Fig.\ \ref{FigClusters}b). The presence of this defect induces a curvature of the structure in agreement with Euler's theorem.  Last, the interaction between the metallic particle and the open tube cap stabilizes it by preventing its closure (Fig.\ \ref{FigClusters}d).

We then calculate the adhesion energies per carbon atom, $E_{a}$,
given by :
\begin{equation} 
\label{eqn:Equation5}
E_{a} =  \frac{(N_{C}+N_{Ni})E_{C/Ni}-N_{C}E_{C}-N_{Ni}E_{Ni}}{N_{C}} \mbox{ ,}
\end{equation} 
where $N_{Ni}$ and $N_{C}$ are the number of Ni and C atoms. $E_{C/Ni}$, $E_{Ni}$ and $E_{C}$ represent the total energies (per atom) of the carbon-covered Ni slab at different surface coverages,
the clean Ni slab and the relevant isolated carbon structure, respectively. They are calculated at 0 K on fully relaxed configurations. In each case, the minimum energy structures are obtained by performing simulated annealing. In this process the initial atomic configuration, initially at high temperature (1000 K),
is slowly cooled down so that the system is at any step at thermodynamic equilibrium. The major advantage of this method is its ability to avoid becoming trapped in local minima. Typical runs
consist in 10$^{3}$ external Monte Carlo loops, each of them randomly performing 10$^{3}$ atomic displacements trials.

\begin{table*}[htbp!]
\caption{Adhesion energies per carbon atom of the carbon structures deposited on the surface of Ni(100).}
\centering
\begin{tabular}{p{0.18\textwidth}p{0.12\textwidth}p{0.12\textwidth}p{0.12\textwidth}p{0.12\textwidth}p{0.12\textwidth}}
\hline
Configurations &Flake &Flake& Flake &Linear chain&Capped tube\\ 
&(Fig.\ \ref{FigClusters}a)&(Fig.\ \ref{FigClusters}b)&(Fig.\ \ref{FigClusters}c)&&(Fig.\ \ref{FigClusters}d)\\
\hline
N$_{C}$& 24 & 54 & 20 & 12 & 60\\
E$_{C}$ (eV/at)& -6.67 &-6.92&-6.55& -5.99   &-6.88\\
E$_{C/Ni}$ (eV/at)& -4.32 &-4.59& -4.30& -4.22 & -4.62 \\
E$_{a}$(eV/at C) & -0.27 &-0.27& -0.58 & -1.01 & -0.29 \\
\hline
\end{tabular}
\label{tab:table1}
\end{table*}

For all the structures tested, the negative value of the adhesion energies, listed in Table \ref{tab:table1}, means that the carbon cluster is stabilized by the surface. For the flakes deposited parallel to the surface, the results clearly show that it is favorable to incorporate pentagons during the initial stages of nucleation. It is therefore tempting to think that the structure containing defects can deform into a dome to better saturate the dangling bonds at their edges.

To investigate further, we analyze the local energies of carbon and nickel atoms in all the previous cases in order to determine which C and Ni atoms atoms gain energy.
\begin{figure}[htbp!]   
\centering
\includegraphics[width=8cm]{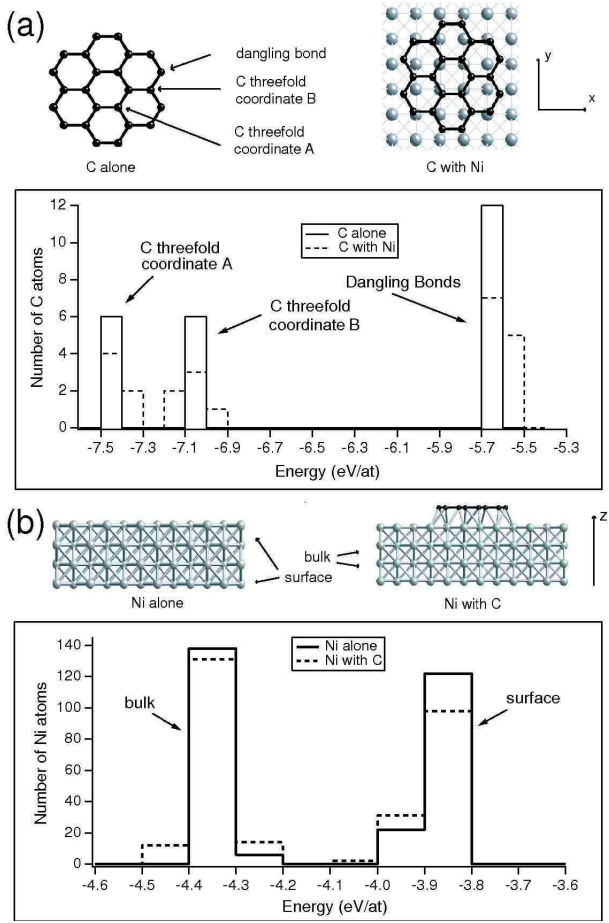}
\caption{Analysis of local energies for a flake of 24 carbon atoms in interaction with a Ni(100) surface. (a) Histogram of C atoms. (b) Histogram of Ni atoms.}
\label{FigHisto}
\end{figure} 
As an example, we plot our results for the flake of 24 atoms in a histogram form in Fig.\ \ref{FigHisto}. After relaxation of the isolated carbon flake (away from the metal surface), we identify three populations of carbon atoms (full lines in Fig.\ \ref{FigHisto}a). First, the three-fold coordinated atoms denoted by A : 6 atoms in a local configuration equivalent to an infinite graphene sheet in which each atom is in a three-fold coordinated state with $sp^{2}$ hybrids forming strong covalent bonds. The energies lie between -7.5 and -7.4 eV/atom, close to the total energy of a perfect graphene sheet calculated with our TB model. Then, the 6 three-fold coordinated atoms located at the periphery of the structure and denoted B: their energies lie between -7.3 and -7.2 eV/atom. The difference between the two previous populations is simply due to the changes in their second neighbor shell. Finally, we have 12 two-fold coordinated atoms at the periphery of the flake, with energies between -5.8 and -5.7 eV/at. These atoms only have two neighbors, so that, in principle, one $sp^{2}$ hybrid is left unbonded, forming a dangling bond. With this analysis, we can estimate the energy of a dangling bond at about 1.80 eV: it is the energy difference, appearing on the histogram, between a C atom located at the surface and a three-fold coordinated A atom. This is close to the 1.78 eV \cite{Bengaard2002} and the 2.26 eV \cite{Fan2003} found in
finite cluster by LDA studies and smaller than the 2.2 to 3.3 eV \cite{Lee1997} found in previous LDA works in the case of a tube. Studying now the flake in contact with the metallic surface, we can
see that the three populations are still present, but modified. In Fig.\ \ref{FigHisto} the histogram still presents three clearly identified regions (dotted lines) equivalent to those described
previously, with a larger distribution due to the accuracy of the relaxation and to the fact carbon atoms that are equivalent in an isolated flake may become no longer equivalent when the flake is
deposited on a surface. The small differences observed however suggest that carbon atoms remains essentially unaffected when deposited on the nickel surface, implying that, in this sense, the interaction between the metallic surface and the carbon structure do not saturate the dangling bonds. This observation is somewhat different from the conclusions drawn by Fan \textit{et al.}, based on \textit{ab initio } calculations, and therefore requires further analysis. In agreement with them, we nevertheless notice that the energy balance presented in Table \ref{tab:table1} reveals a weak adhesion energy, meaning that the flake is more stable on the surface than free standing.  Since this stabilization does not come from the carbon atoms, these results imply that nickel atoms should be stabilized by the presence of carbon.
\begin{figure}[htbp!]  
\centering
\includegraphics[width=8cm]{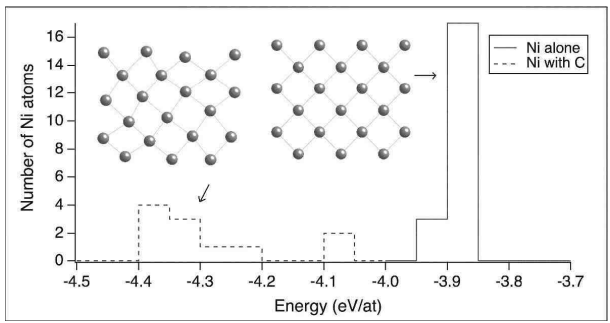}
\caption{Histogram of Ni atoms close to the flake of graphene. Reconstruction of the Ni(100) surface when C atoms are present.}
\label{FigHistoNi}
\end{figure} 
We confirm this with the analysis of the local energies of Ni atoms shown in Fig.\ \ref{FigHisto}b and Fig.\ \ref{FigHistoNi}. In the case of the bare relaxed Ni slab, we observe two different populations (full lines in Fig.\ \ref{FigHisto}b). The first corresponds to the bulk atoms which have 12 neighbors and energies
between -4.4 and -4.3 eV/atom. The total energy of a face centered cubic bulk Ni is $E_{tot}^{i} = $-4.44 eV/atom  in our TB model. The second population corresponds to the 8-fold coordinated surface atoms. These are less stable than atoms in bulk and have energies between -4.0 and -3.8 eV/atom. In the presence of the carbon structure, both populations are still present. It is important to note that 288 atoms of Ni have been reported on this histogram. Thus, the difference observed on Fig.\ \ref{FigHisto}b which seems weak when compared to the analysis of C atoms where only 24 atoms appear, corresponds to a larger contribution to the total energy. The quantitative analysis of the local energies shown in Fig.\ \ref{FigHistoNi}, demonstrates that the nickel atoms, located at the surface and in direct interaction with the carbon structure gain energy. On the one hand, the average total energy of nickel atoms at the surface and far away from carbon atoms is around -3.90 eV/atom. This value is similar to the one obtained without carbon atoms (full lines in Fig.\ \ref{FigHisto}b); on the other hand, the nickel atoms close to the carbon structure are stabilized (Fig.\ \ref{FigHistoNi}). The energy gain is approximatively equal to 0.4 eV/atom and is associated with a slight reconstruction of the Ni surface underlying the flake, as shown in Fig.\  \ref{FigHistoNi}. In the case of the other flat structure containing 54 carbon atoms, the local energy analysis leads to the same conclusions: the local energies of carbon atoms are not modified but surface nickel atoms below the flake are stabilized with roughly the same energy gain. 

\begin{figure}[htbp!]   
\begin{center}
\includegraphics[width=8cm]{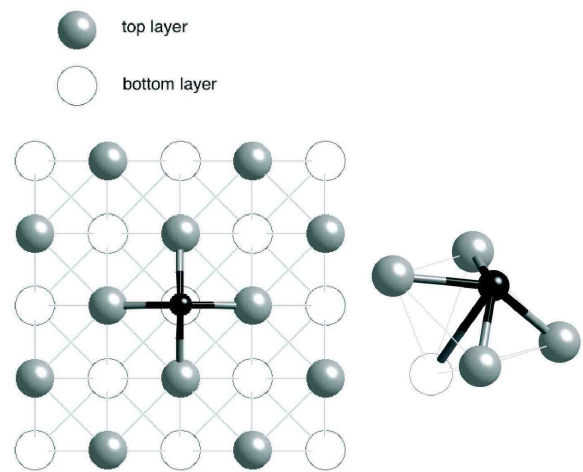}
\end{center}
\caption{Illustration of a C atom occupying a hollow site of Ni(100) surface.}
\label{FigHollow}
\end{figure}

The same energy balance is obtained for the flake containing a pentagon.  For the linear chain, the one-fold coordinated, end of chain, carbon atoms interact strongly with the surface and penetrate into the metallic substrate to occupy a semi-octahedral site of the (100) Ni structure (Fig.\ \ref{FigHollow}) with an energy gain of 0.5 eV. The same effect appears with the tube embryo where some carbon atoms are located in the fourfold hollow sites. These results show clearly that the anchoring of carbon atoms to the catalyst is energetically favorable during the nucleation process. They gain energy by occupying the semi-octahedral sites of the fcc nickel (100) surface.

Finally the adhesion process of carbon sheets on Ni (100) is slightly more complex than anticipated. The adhesion energy of flat sheets is mainly due to the energy gain of the nickel atoms below these sheets. When they curve to form caps, the energy gain becomes concentrated on the carbon and nickel atoms atoms close to their edge. In this case one might argue that dangling bonds are saturated, but finally, this is a weak effect which does not play very much in favour of curved caps: the energy of adhesion of flat and curved sheets are similar and small compared to dangling bond energies, for small clusters at least. 

At this point it is useful to stress that it can be difficult to choose between various decompositions and normalizations of energies of formation or adhesion. Depending on the point of view, it can be fruitful to count energies per carbon atom of the whole structure (flat sheet) or per edge atom (curved sheet). Our model has the advantage to provide us with unambiguous local (site) energies. Bond energy analyses could be performed as well but would be more complex to use in the present context.

\section{Discussion} 

Our results allow us more generally to understand the mechanisms by which carbon atoms adsorb on the Ni(100) surface. Experimentally, investigations including low-energy electron diffraction \cite{Onuferko1978}, surface extended-x-ray-absorption fine structure \cite{Bader1987}, and scanning tunneling microscopy \cite{Klink1993} have provided a large amount of information on the
interaction of C with Ni(100). When carbon adsorbs on Ni(100) with a coverage smaller than one third of a monolayer (ML), it occupies hollow sites and does not change the symmetry of the outermost metal
layers (Fig.\ \ref{FigHollow}).  This situation results in strong interactions (see table \ref{tab:table2}) and corresponds to the incorporation of isolated or one-fold coordinated C atoms in the hollow sites, observed during our calculations.

\begin{table*}[htbp!]
\caption{Binding energies (E$_{b}$) of a C atom on a Ni(100) slab. R is the distance between C and Ni atom.}
\label{tab:table2}
\begin{minipage}[htbp]{15cm}
\begin{tabular}{p{0.005\textwidth}p{0.16\textwidth}p{0.20\textwidth}p{0.17\textwidth}}
\\
\hline
&&E$_{b}$ (eV)& R (\AA)  \\
\hline
&Theory \footnotemark[1]& -7.43, -6.50, -8.10, & 1.91, 1.85, 1.77\\
&&-7.69, -8.43&\\
&Experiment \footnotemark[1]& -7.35, -7.37, -7.55 & 1.80, 1.82, 1.89\\
&Present work&  -8.21 & 1.90 \\
\hline
\end{tabular}
\footnotetext[1]{From Ref. \cite{Zhang2004} and references cited therein. }
\end{minipage}
\end{table*}

As the coverage exceeds 0.33 ML, the surface undergoes a reconstruction known as the ``clock reconstruction" described previously (Fig.\ \ref{FigClock}). This reconstruction, which has also been
studied theoretically \cite{Alfe1999,Stolbov2005}, is indeed observed during our MC calculations (Fig.\ \ref{FigHistoNi}). 

The adhesive properties of carbon clusters on Ni(100) surface then result from a competition between the formation of Ni-C and C-C bonds. For structures containing few C-C bonds, the competition between a carbide and a demixed system favors the formation of Ni-C bonds and the C atoms occupy the hollow sites of Ni(100) present at the surface. For planar flakes, C-C interactions predominate and, as described above, the weak energy of adhesion is principally related to the energy gain of the (reconstructed) metallic surface below the flakes. Tube embryos have an intermediate behaviour: C-C interactions predominate on the caps and Ni-C anchoring bonds predominate at their edges.

It is interesting to compare briefly our results with those obtained for the (111) surface. At low carbon concentration the situation is similar, with predominant Ni-C interactions, even if, due to the different topology of the surfaces, the preferred position for the (111) surface for carbon is a subsurface octahedral site.  At higher concentration we expect the formation of a graphene sheet in quasi perfect epitaxial relationship with the Ni substrate and a vanishingly small energy of adhesion \cite{Amara2}. We have also calculated the energy of adhesion of a small flat cluster made of 24 atoms. As in the case of the (100) surface we find a weak energy of adhesion equal to -0.26 eV/C atom, and the local energy analyses show that the energy gain is concentrated on the edge atoms, without any important saturation of the dangling bonds. In their analysis of different caps on Ni (111) Reich \textit{et al.} have also noticed that the relative energies involved are also fairly weak \cite{Reich2005}.

\section{Conclusion} 

In summary, we have presented tight-binding Monte Carlo simulations of the interaction of carbon structures with Ni(100) surfaces Our calculations show that during the first stages of the nucleation the incorporation of defects and the anchoring of carbon atoms to the surface are energetically favorable. Even in the presence of a metal substrate, the carbon dangling bonds parallel to the surface still exist and the metal-carbon system gains energy by reconstructing the underlying metallic surface, close to the well-known clock reconstruction. For curved structures, the formation Ni-C bonds is more efficient, and C atoms prefer to be located in semi-octahedral sites of Ni present at the surface.\\

Part of this work has been supported by the Belgian Program on Interuniversity Attraction Poles (PAI5/1/1) on ``Quantum Size Effects in Nanostructured Materials''.


\begin{thebibliography}{99}
\bibitem{note} Unfortunately it is difficult to reproduce the analysis from the data presented in the article.
\end{thebibliography}
\end{document}